\documentclass[a4paper, onecolumn, twoside, 12pt]{article}
\usepackage[utf8]{inputenc}
\usepackage[greek,english]{babel}
\usepackage{graphicx}
\usepackage{verbatim}
\usepackage{wrapfig}
\usepackage{subfig}
\usepackage{float}
\usepackage{amsmath,amssymb}
\usepackage{array}
\usepackage{authblk}
\usepackage[font=small,labelfont=bf]{caption}

\date{}                     
\title{\bf Cosmic Ray RF detection with the ASTRONEU array}

\author[1]{I. Manthos\footnote{corresponding author: i.manthos@fme.aegean.gr}}
\author[1]{I. Gkialas}
\author[2]{G. Bourlis}
\author[2]{A. Leisos}
\author[2]{A. Papaikonomou}
\author[2]{A.G. Tsirigotis}
\author[2,3]{S.E. Tzamarias}
\affil[1]{Department of Financial and Management Engineering, University of the Aegean, Chios, Greece}
\affil[2]{Physics Laboratory, School of Science and Technology, Hellenic Open University, Patras, Greece}
\affil[3]{Department of Physics, Aristotle University of Thessaloniki, Thessaloniki, Greece}

\begin{document}
   \maketitle

\begin{abstract}
Results will be shown from the ASTRONEU array developed and operated in the outskirts of Patras, Greece. An array of 9 scintillator detectors and 3 antennas were deployed to study Extensive Air Showers (EAS) as a tool for calibrating an underwater neutrino telescope, possible other applications in muon tomography, education purposes, and last but not least, the detection of air showers via their electromagnetic signature. This is the first stage of a total of 24 scintillator counters and 6 RF antennas to complete the array. In this work, results with regard to the electromagnetic detection of showers will be shown. The method of operation and analysis will be presented. The purpose of this project was to demonstrate the adequacy of the method to detect cosmic events even in the presence of high urban electromagnetic background, using noise filters,  timing, signal polarization, and eventual comparison with well understood event reconstruction using the scintillator detectors.  The results indicate that cosmic showers were detected and the method can be used for the complete array.
\end{abstract}

\section{Introduction}

The cosmic ray showers are traditionally detected using large arrays of particle detectors. In the sixties, a new, complementary method of detecting cosmic ray showers was proposed, using the RF signature of a cosmic ray event. This mechanism \cite{Askaryan_radio_emission}, exploits the electromagnetic signal emitted by the electromagnetic interactions of the shower charged particles. The phenomenon was detected for the first time in 1965 \cite{Askaryan_radio_emission}. The interest in this method of detection soon subsided because of inadequate technology. In the beginning of the 21\textsuperscript{st} century, the interest was renewed given the advances in fast digital electronics and the low cost of the antennas compared to the scintillator counter detectors. Since then, a number of experiments have studied this method, \cite{Lofar, Codalema} and more are planned for the future. The advantage of this method of detection is its large duty cycle, of the order of 95\% compared to other methods, such as fluoresence and Cherenkov. Two proposed  mechanisms to explain the phenomenon are the geomagnetic emission mechanism \cite{Scholten}, and the charge excess mechanism \cite{Askaryan_charge_excess}.\\
An effort was made in Greece in the past years to develop the scintillator array HEllenic LYceum Cosmic Observatories Network (HELYCON) \cite{IDM, HELYCON} for educational and underwater neutrino telescope (NESTOR \cite{Nestor}, KM3NeT \cite{Km3net}, ANTARES \cite{Antares}) related purposes.  In the most recent application of these concepts, an improved combined scintillator detector and RF antenna array, ASTRONEU,  was developed in the campus of the Hellenic Open University in the outskirts of the city of Patras, Greece.\\
In this paper the first results on RF detection of cosmic ray induced EAS using the initial phase of the ASTRONEU  array will be presented. In section 2, a brief description of the theoretical background is given. In section 3, the deployed experimental setup will be described. Then, in the following section 4, the event selection procedures are presented. The main analysis and the results are presented in section 5 and, finally, some conclusions are drawn in section 6. 

\section{Theory}
The main sources of the RF are the emission due to geomagnetic and  charge excess (Askaryan effect) \cite{Huege} mechanisms. The geomagnetic emission is a result of the acceleration of $e^-$ and $e^+$ in opposite directions under the influence of the  Lorentz force. Consequently, a time varying transverse current is developed in the evolution of the shower, depending on the varying number of the charged particles of the shower. The electric field produced by this current is linearly polarized along an axis perpendicular to the shower propagation axis and is independent of the observer's position. The charge excess emission mechanism is the secondary contribution of the emitted signal (10\% to 20\% of the total emission). During the evolution of the EAS a negative charge excess occurs in the shower front, due to the annihilation of the positrons in contrast to the electrons. In addition there is an enrichment of the shower front with  electrons from the air molecules.  Moreover, the formed positive ions are following with smaller velocities the shower, presenting positive charge excess in the shower tail. In this way, an electric field is created along the shower axis, contributing to the detected signal. Its polarization vector arises from the projection of the electric field onto the observation plane. The detected RF signal arises from the superposition of the two emission mechanisms when their electric fields add constructively. The tip of the polarization vector moves in an ellipse resulting from the mixture of the linear (geomagnetic) and circular (charge excess) polarizations. The relativistic velocity of the particles make the emission strongly directional (Cherenkov – like) restricting the detection area. 

\section{Experimental setup}

The ASTRONEU array reported in this article consisted of three stations, with each station comprising 3 scintillator counters and 1 Codalema type RF antenna \cite{Codalema, Astroneu} which samples the RF waveform at $1\,GHz$. The 3 scintillator counters form an approximate equilateral triangle in stations 1 and 3 while the site constraints dictated that the station 2 formed an amblygonal triangle. The scintillator detectors were readout with the use of the Quarknet \cite{qnet} electronics which implement the Time over Threshold (ToT) technique.  The station trigger was generated when all three scintillator detectors of the station had signal above the $9.7\,mV$ threshold corresponding to roughly 2 Minimum Ionizing Particles within the trigger time window of $120\,ns$, corresponding to maximum distance between any two scintillator counters in a station. The absolute station trigger time is defined as the Quarknet GPS time of the 3\textsuperscript{rd} (last) signal. This trigger signal was fed into the antenna external trigger input. The antenna upon receiving the external trigger recorded the full 2560 buffer samples. The antenna trigger time was set in such a way so that any cosmic related signal would be recorded near the center of the buffer. \\
The ASTRONEU cosmic ray telescope is deployed in an urban environment with an abundance of man made RF signals. For other experiments aiming to detect the RF signal of the EAS, such as Codalema \cite{Codalema}, LOFAR \cite{Lofar} and AERA \cite{Aera}, minimizing the man made noise was of the utmost importance. In the case of ASTRONEU, which it should be noted is a pilot project, a trial DAQ self-trigger mode was used. However, even with the increase of the threshold value, there was a constant saturation in the trigger rate at $28\,Hz$. This fact would increase the value of the primary particle energy threshold and also make the distinction of the cosmic events among this ``sea" of noise very difficult. Taking advantage of the hybrid character of the telescope and the detection of cosmic events from the particle detectors, the antenna of each station operates in external trigger mode with the trigger coming from the digitization card of the scintillator signal, as described above. \\
This DAQ method of external triggering, besides not being autonomous, has also the disadvantage that the antenna efficiency is constrained by  the particle detector station efficiency. However, the self-trigger DAQ mode has not been efficiently used  up to now in any experiment in the world, while the only real low noise site is Antarctica as revealed from RASTA \cite{Rasta} and ANITA \cite{Anita}. On the other hand, the hybrid operation of antennas and particle detectors can be considered as an asset, ensuring that the detected event is of a cosmic origin and the RF signal analysis can reveal information for the EAS, hard to extract from the particle detection, most importantly the primary particle energy.

\section{Event selection}

During the data analysis presented in \cite{Astroneu2}, a total of  more than 600000 events from all the three HELYCON stations, fulfilling the quality criteria and accepted as EAS events were collected in the period August 1, 2014, to March 16, 2016. Of course, the majority of these events comes from primaries with energies below $10^{17}eV$,  where the antenna efficiency is very low. Events with two hit antennas were analysed,  meaning stations 1 - 2 and 1 - 3 as the distance between stations 2 - 3 makes the existence of cosmic origin signal in both antennas very improbable, as the RF signal of the EAS is detected as far as $120\,m$ from the shower axis, and the inclined EAS that have larger footprint are less efficiency detected from the horizontal particle detectors used for the antenna triggering. For the same reason there are not triple coincidence events. The selection of EAS events is done forming a double coincidence between triggered stations, using the station trigger absolute times, in a time window of $1500\,ns$.\\
The event search for double coincidences between stations 1 - 2, yielded 1354 events on station 1 antenna and 1292 events on station 2 antenna. The difference in the number of events is due to small time intervals that the antenna of station 2 remained inactive for technical reasons.

\section{RF signal analysis}

\begin{figure}[h]
\includegraphics[width=8cm]{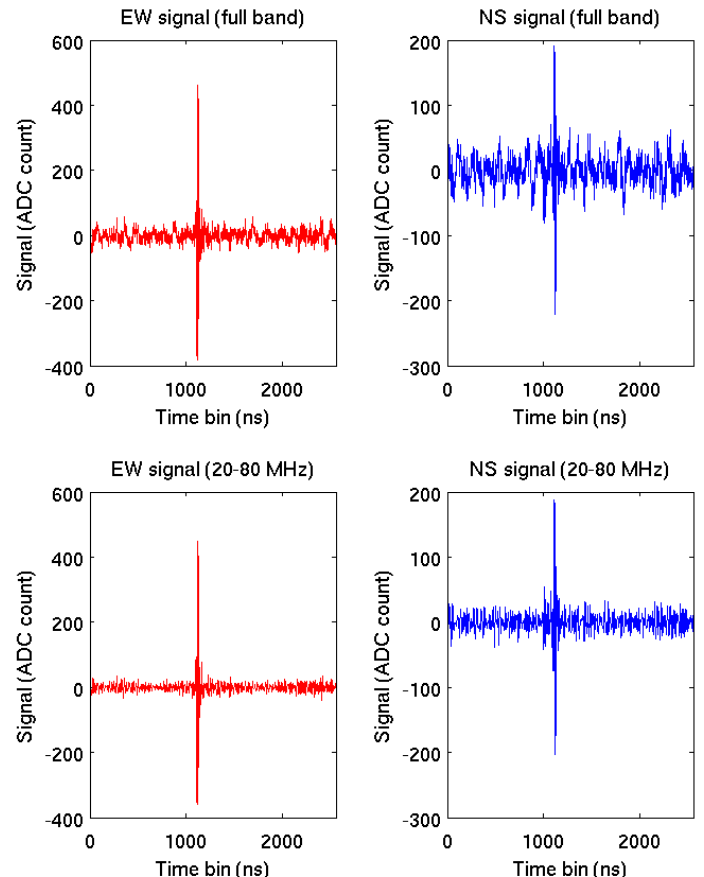}
\centering
\caption{\textbf{Above:} Raw signal and \textbf{below:} Filtered signal, for each antenna pole for a characteristic event.}
\label{fig:filter}
\end{figure}

The recorded antenna signals were analysed following the methodology developed by the Codalema experiment, with some of the filtering code having been developed by Codalema.\\
For the signal analysis filtering, it is essential to keep only the frequencies between $20 - 80\,MHz$. This is necessary because for frequencies below $20\,MHz$ the ionosphere is strongly reflective, so that man made signals from long distances can still contribute to the noise. At high frequencies, over $80\,MHz$, the presence of strong signals from radio FM band necessitates the exclusion of this range from the search for cosmic events. The filtering procedure is implemented in two phases. Initially, the waveform is subject to Tukey filtering, suppressing the outliers, and then, a Fast Fourier Transform (FFT) is applied. After this transformation, the non-desirable frequencies are removed and subsequently the inverse Fourier Transform (FFT\textsuperscript{-1}) is applied, returning the filtered signal for further analysis. Typical example of this procedure is shown in Figure \ref{fig:filter}, for the two poles of the antenna; the upper part depicts the raw signals as recorded for the full waveform ($2560\,ns$). In the lower part of the image the filtered, clean signals are shown. 

\subsection{Characterization criteria for cosmic events and application to events recorded by stations 1 and 2}
\begin{figure}[h]
\includegraphics[width=14cm]{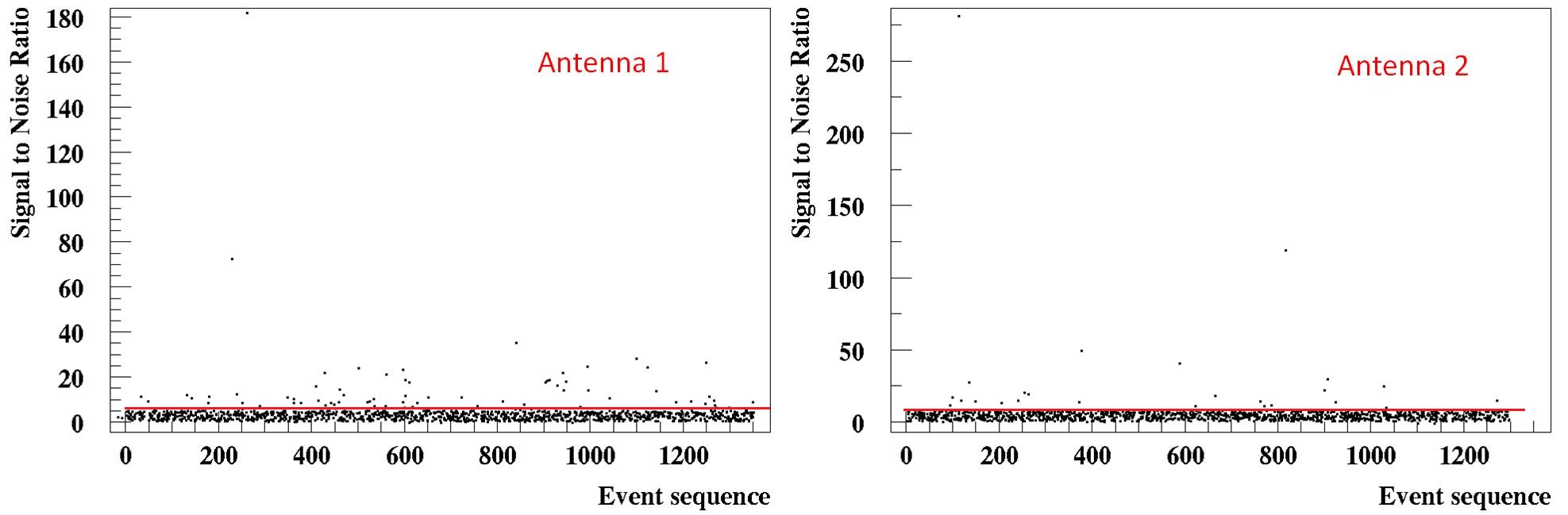}
\centering
\caption{Signal to Noise Ratio for each double coincidence event between stations 1 and 2 for  antennas 1 \textbf{(left)} and 2 \textbf{(right)}.}
\label{fig:snr}
\end{figure}

\begin{figure}[h]
\includegraphics[width=14cm]{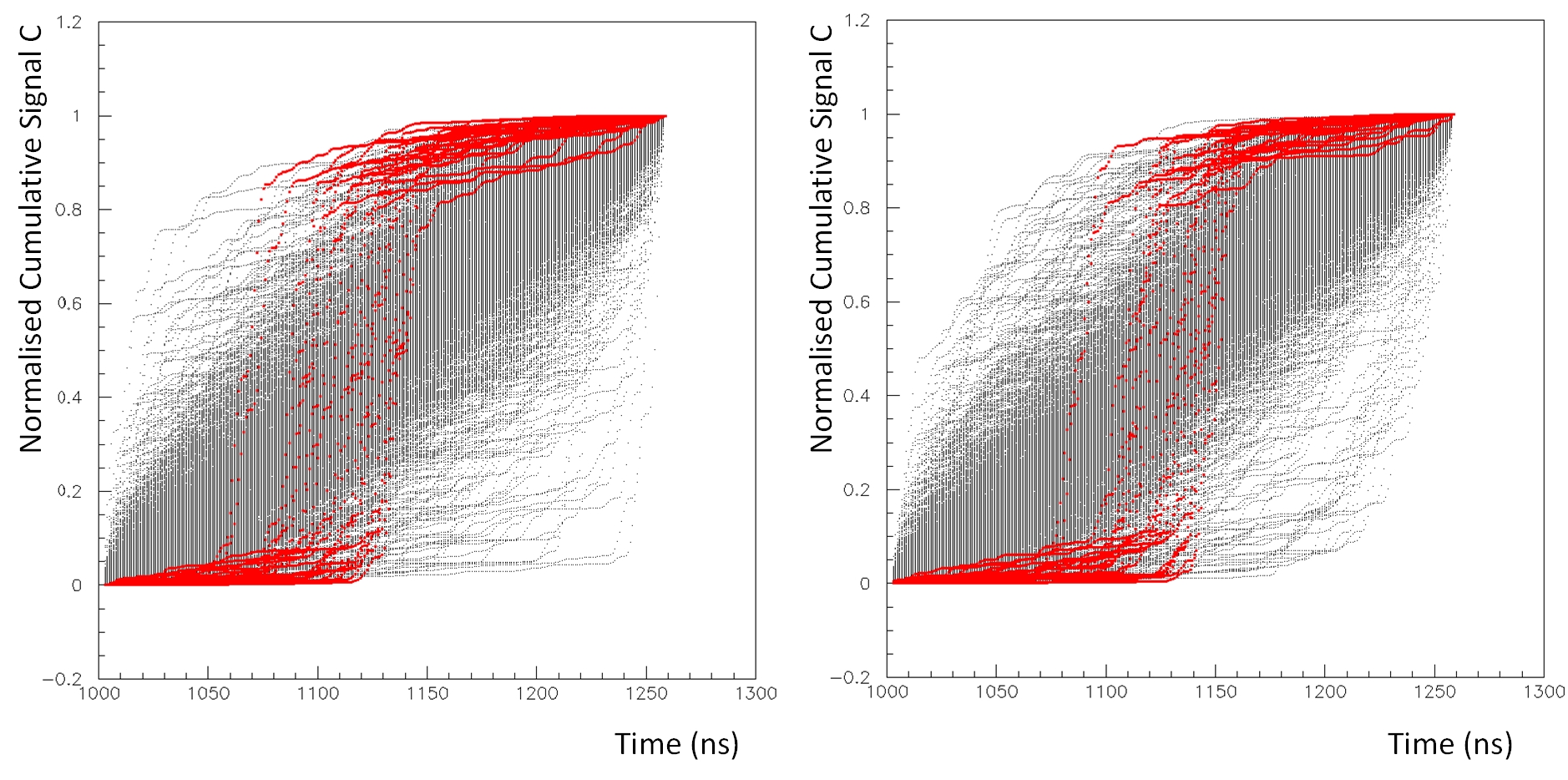}
\centering
\caption{Cumulative signal in a $\pm 128\,ns$ window around the peak voltage for antenna 1 \textbf{(left)} and antenna 2 \textbf{(right)}. Cosmic candidate events, with rise time $\leq 28\,ns$ are depicted  in red color.}
\label{fig:cumulative}
\end{figure}

\begin{figure}[h]
\includegraphics[width=14cm]{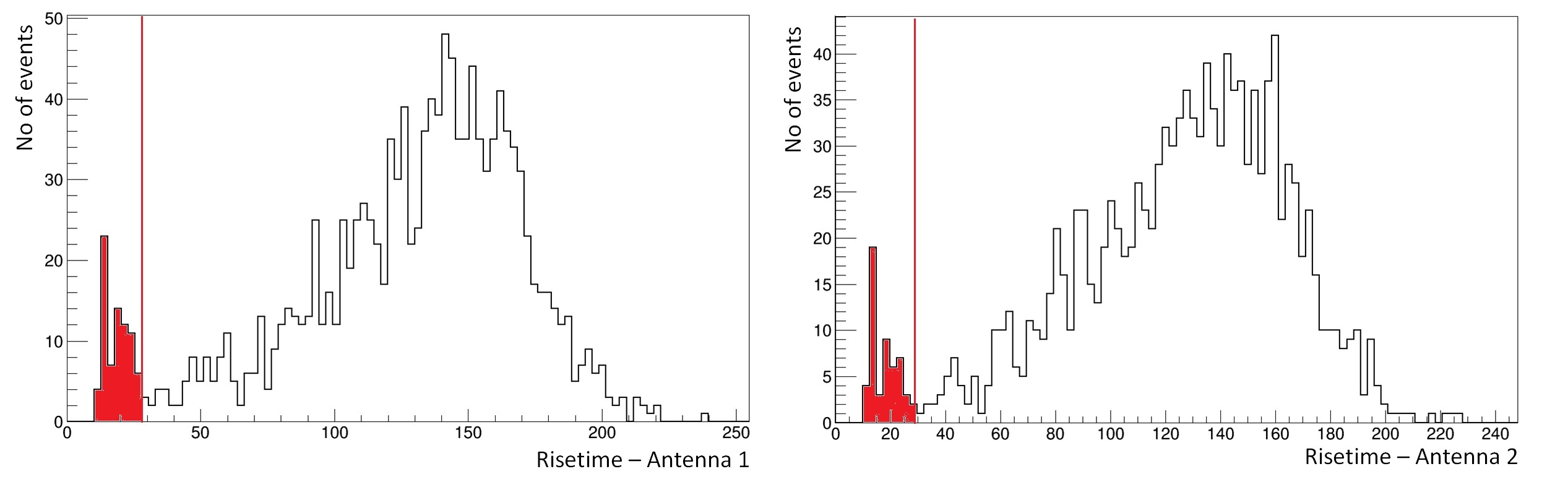}
\centering
\caption{Distribution of transition time from 10\% to 70\% of the total cumulative signal in a $\pm 128\,ns$ window around the peak voltage for antenna 1 \textbf{(left)} and antenna 2 \textbf{(right)}.}
\label{fig:risetime}
\end{figure}

\begin{figure}[h]
\includegraphics[width=14cm]{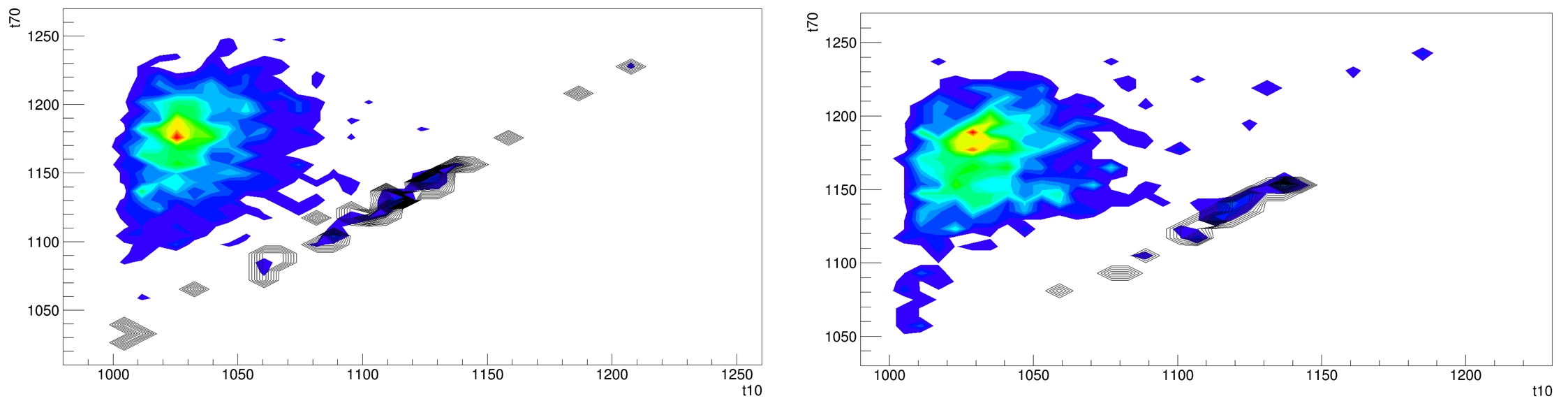}
\centering
\caption{Plot of 70\% cumulative signal time versus 10\% cumulative signal time for antenna 1 \textbf{(left)} and antenna 2 \textbf{(right)}.}
\label{fig:risetime2}
\end{figure}

\begin{figure}[h]
\includegraphics[width=14cm]{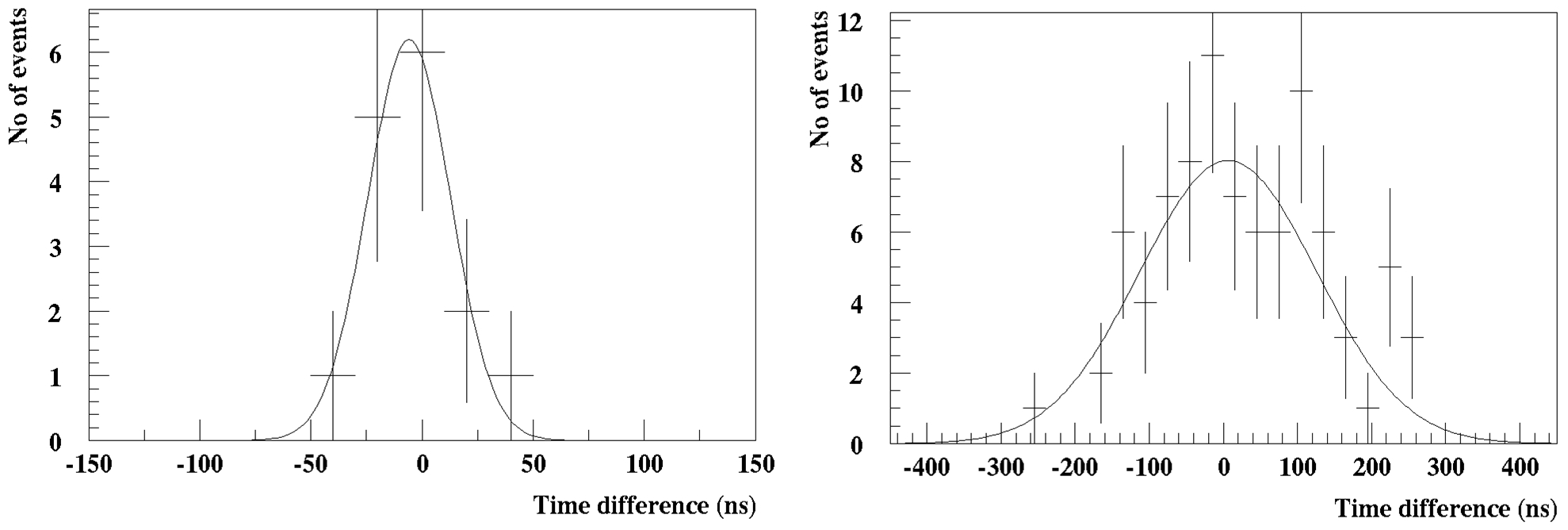}
\centering
\caption{Distribution of the recorded time difference between antennas 1 and 2 and the calculated time difference from the reconstructed angles for the RF simultaneous events passing all 3 criteria (15 events), at the \textbf{left}, and for events presenting signal rise time $>28\,ns$ and not passing the $SNR$ criterion, at the \textbf{right}.}
\label{fig:recos_ant1}
\end{figure}

An event is characterized as being of cosmic origin, when three specific criteria for the signal are fulfilled:
\begin{itemize}
\item It presents an intense peak localized in a narrow time interval.
\item It exhibits short rise time.
\item Its polarization is approximately linear (EW vs NS).
\end{itemize}
With regard to the first criterion, having in mind that the thickness of the shower front is of the order of about $10\,m$,  a time duration of about $30\,ns$ is expected. Knowing that the majority of the particles are  in the shower front, a sudden, strong pulse followed from a small number of pulses with decreasing amplitude is expected. When the scintillation detectors signals have formed a trigger, the electromagnetic waveform contained in the buffer is recorded, in particular a total of $2560$ voltage values. The expected position of the signal in the buffer is well defined. Considering the external triggering of the antenna, delays are caused by the signal transmission cables and  the Quarknet card response. So, the antenna software was adjusted so that  the signal appears around the center of the buffer, around element $1100$ (out of a total of $2560$ elements) of the saved waveform. Such a signal is shown in Figure \ref{fig:filter}. To analyse the antenna recorded  events, the EW component was used due to the greater possibility of detection in this direction. This is explained considering the effect of the Lorentz force on the charged particles of the EAS, as the vector of the magnetic field points to the South, reinforcing the possibility of detection in the perpendicular direction EW. This discussion is quantified calculating the Signal to Noise Ratio ($SNR$). 
\begin{figure}[h]
\includegraphics[width=7cm]{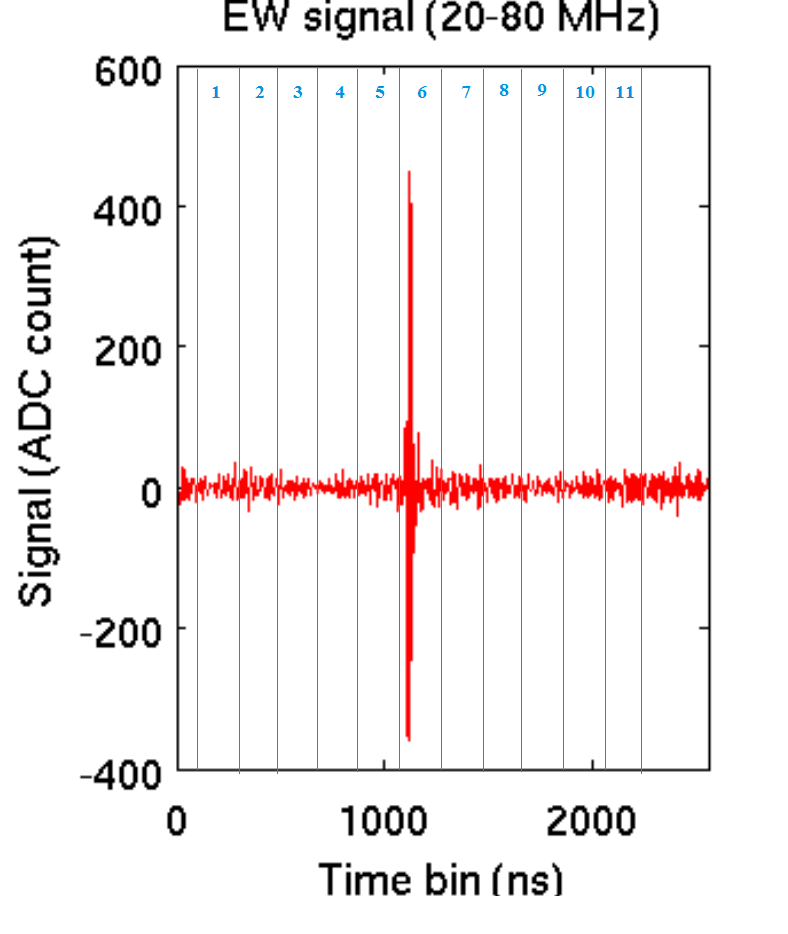}
\centering
\caption{Binning of the complete waveform in $11$, $200\,ns$ windows for $SNR$ calculation.}
\label{fig:binning}
\end{figure}
As depicted in Figure \ref{fig:binning} most of the $2560\,ns$ window is partitioned in $11$ time windows with duration $n=200\,ns$ each. Starting from $t=50\,ns$  the quantity
\begin{equation}
E_{i}={1\over n}\sum_{k=(i-1)\cdot n+51}^{i\cdot n +50}V\left( k\right) ^{2}
\label{eq:Ei}
\end{equation}
was calculated for each window ($i$ running from $1$ to $11$). The 6\textsuperscript{th} window, from $1051\,ns$ to $1250\,ns$, contains the signal. The remaining $10$ windows contain the background noise. The signal over background ratio for each event was calculated using the following formula:
\begin{equation}
SNR=\dfrac{E_{6}}{\frac{1}{10} \sum_{i=1, i\neq 6}^{11}E_{i}}
\end{equation}
As shown in Figure \ref{fig:snr}  plots,  the majority of the simultaneously recorded events (by the particle detectors) between stations 1 and 2,  do not exhibit any significant signal standing out in the window of interest, something expected as most of the events derive from primaries below $10^{17}\,eV$ where the RF signal is not detected. The events above the red line meet the first criterion and are considered candidate events of cosmic origin. The criterion for antenna 1 is set to $SNR > 5$, while for antenna 2 it is set to $SNR > 8$.
Useful conclusions can be revealed from the mean values of the squared voltages in each window given by Equation \eqref{eq:Ei}, as tabulated in Tables \ref{tab:noise1} and \ref{tab:noise2} for the events fulfilling the $SNR$ criterion. It is observed  that both antennas record the same noise level in the EW pole, in windows from 1 to 5, and 7 to 11, presenting small deviations. However, the value $E_6$  in each antenna is much larger than the rest of the values, as expected, since it contains the signal.\\
\begin{table}[h]
\begin{center}
\begin{tabular}{>{\centering\arraybackslash}p{1.3cm}|>{\centering\arraybackslash} p{0.7cm}>{\centering\arraybackslash} p{0.7cm}>{\centering\arraybackslash} p{0.7cm}>{\centering\arraybackslash} p{0.7cm}>{\centering\arraybackslash}p{0.7cm}>{\centering\arraybackslash} p{0.7cm}>{\centering\arraybackslash} p{0.7cm}>{\centering\arraybackslash} p{0.7cm}>{\centering\arraybackslash} p{0.7cm}>{\centering\arraybackslash} p{0.7cm}>{\centering\arraybackslash} p{0.7cm}}
Window & 1 & 2 & 3 & 4 & 5 & \textbf{6} & 7 & 8 & 9 & 10 & 11  \\ 
\hline 
$E_i (V^{2})$ & 0.158 & 0.122 & 0.132 & 0.114 & 0.169 & \textbf{1.916} & 0.226 & 0.131 & 0.140 & 0.098 & 0.123  \\  [0.35cm]
\end{tabular}
\caption{Mean value of squared Voltage, as measured from antenna 1 from cosmic candidate events (after $SNR$ criterion), according to the binning of Figure \ref{fig:binning}.}
\label{tab:noise1}
\end{center}
\end{table} 
\begin{table}[h]
\begin{center}
\begin{tabular}{>{\centering\arraybackslash}p{1.3cm}|>{\centering\arraybackslash} p{0.7cm}>{\centering\arraybackslash} p{0.7cm}>{\centering\arraybackslash} p{0.7cm}>{\centering\arraybackslash} p{0.7cm}>{\centering\arraybackslash}p{0.7cm}>{\centering\arraybackslash} p{0.7cm}>{\centering\arraybackslash} p{0.7cm}>{\centering\arraybackslash} p{0.7cm}>{\centering\arraybackslash} p{0.85cm}>{\centering\arraybackslash} p{0.7cm}>{\centering\arraybackslash} p{0.7cm}}
Window & 1 & 2 & 3 & 4 & 5 & \textbf{6} & 7 & 8 & 9 & 10 & 11  \\ 
\hline 
$E_i (V^{2})$ & 0.148 & 0.137 & 0.134 & 0.166 & 0.135 & \textbf{3.89} & 0.175 & 0.140 & 0.218 & 0.144 & 0.113  \\  [0.35cm]
\end{tabular} 
\caption{Mean value of squared Voltage, as measured from antenna 2 from cosmic candidate events (after $SNR$ criterion), according to the binning of Figure \ref{fig:binning}.}
\label{tab:noise2}
\end{center}
\end{table}
The distances among the particle detectors of station 2 are greater, resulting in EAS with larger particle numbers being detected. The trigger threshold value for each detector, corresponding to a minimum number of particles incident on the detector, are equal for all detectors. For station 2, then, this is achieved either by EAS of higher energy or by EAS with their impact point closer to station 2. 
\begin{figure}[h]
\includegraphics[width=14cm]{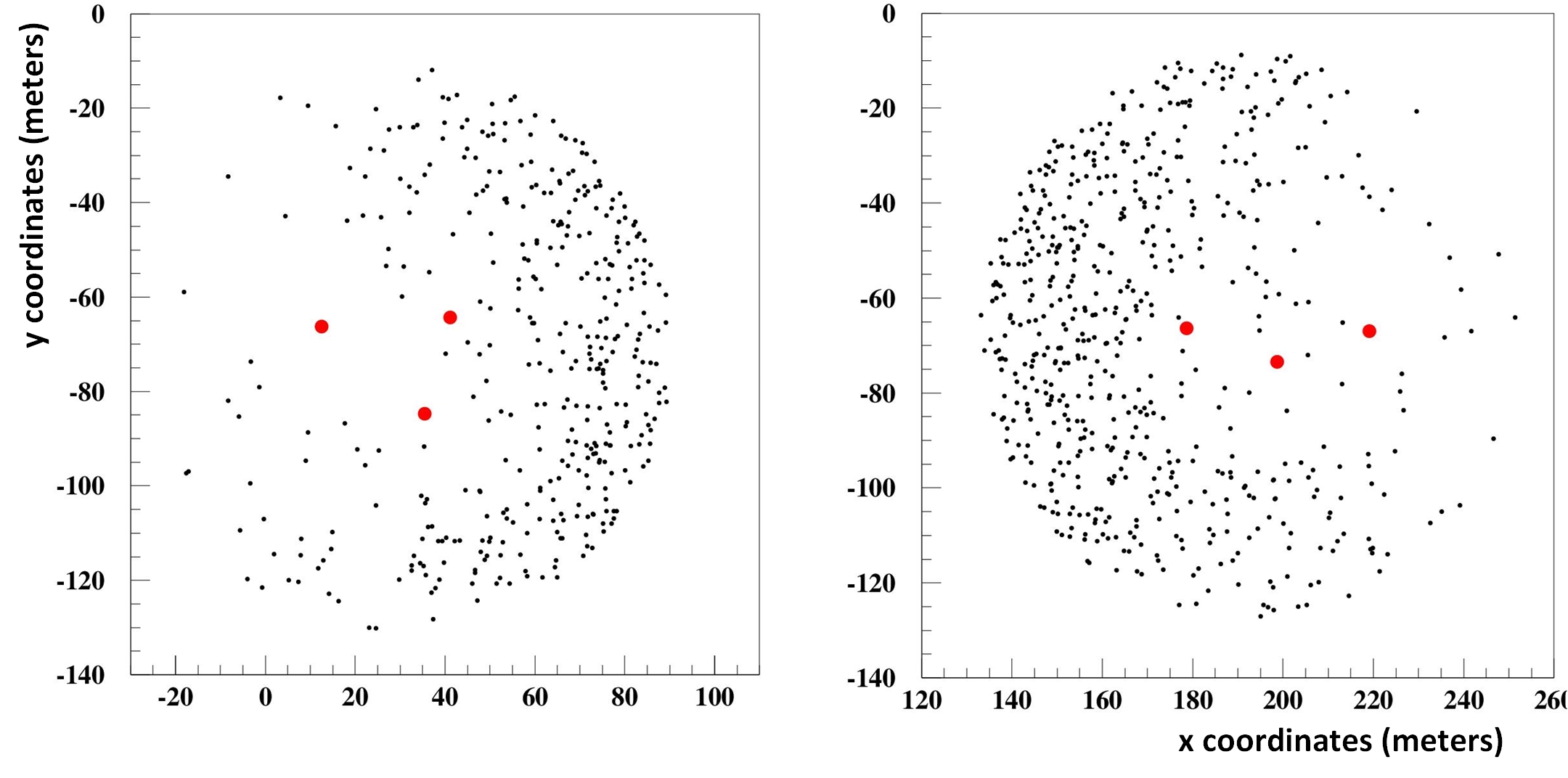}
\centering
\caption{EAS impact point coordinates with respect to the scintillator counters (red dots) with primary particle energy $>10^{17}\,eV$ triggering simultaneously stations 1 and 2, within a circle of $60\,m$ radius from the center of station 1 \textbf{(left)} and station 2 \textbf{(right)} as predicted by the simulation software.}
\label{fig:corshits}
\end{figure}
Consequently, for EAS triggering simultaneously stations 1 and 2, a shift of the hit point on the ground from the geometrical center, closer to station 2, will be observed as it is demonstrated by the simulation software, involving Corsika \cite{Corsika_manual} and HOURS \cite{hours} packages,  as shown in Figure \ref{fig:corshits}. In a total of $17786$ simulated events, corresponding to an equivalent lifetime approximately 10 times more that the actual data taking period analysed in this work, $388$ events are detected within a circle of $60\,m$ radius from the barycentre of station 1, while $589$ events are detected within the same size circle centred at the barycentre of station 2. In general, by studying the signals one by one, it is observed that station 2 has detected less cosmic candidate signals, but with better characteristics, being more ``clear" compared to those detected by station 1. Thus, the influence of the particle detectors geometry on the quality of the detected RF signal, is evident.\\
The second criterion takes advantage of the fact that the signal should be of the order of $30\,ns$, consistent with the $10\,m$ shower front thickness. To take into account the time duration of a cosmic generated signal, it was proposed to study the cumulative signal over a large time window and see if there is a difference in rise time for different types of events. After finding the peak voltage point of the waveform, the sum of the squares of the measured voltages in each bin (each bin corresponds to $1\,ns$) in an area $\pm 128\,ns$ around the peak was calculated and normalized according to Eq. \eqref{eq:norcumsig}
\begin{equation}\label{eq:norcumsig}
C_{\left( i\right) }={{\sum_{k=max-128}^{max-128+i}s\left( k\right)^{2}} \over {\sum_{k=max-128}^{max+128}s\left( k\right)^{2}}}
\end{equation}
where max corresponds to the bin with the peak voltage in the waveform. By plotting the cumulative signal for each event, it can be seen  that some events demonstrate different behaviour, concentrating large percentage of the signal in a short time interval, as shown in Figure \ref{fig:cumulative}, for the coincidence events  between stations 1 and 2. 
The time interval needed for the cumulative signal to reach from 10\% to 70\% of the total is plotted in Figure \ref{fig:risetime}. Events with fast rise time seem to form a separate structure in the low end of the distribution, indicating maybe a different, cosmic, origin of these events, in contrast to the bulk of the events with higher rise times, which can be attributed to background events. The cut value for this separation is set to $28\,ns$ and denoted with the red vertical line. The cosmic candidate events are also set in red color in both Figures \ref{fig:cumulative} and \ref{fig:risetime}.
Another way of showing this effect is by plotting  the 70\% of the cumulative signal time  against the 10\% of the cumulative signal time. The result is presented in Figure \ref{fig:risetime2}, with the events lying in the straight line to be the cosmic candidate events, depicted in red color in Figures \ref{fig:cumulative} and \ref{fig:risetime}. 
The third criterion is relevant with the polarization of the signal, which was determined by plotting  the signal  detected by the EW pole versus the signal detected by the NS pole of the antenna, for each $1\,ns$ bin. The expected polarization of a cosmic event, is expected to form an ellipse, as mentioned in a previous section. Finally, the polarization check for the events that already have successfully passed the first two criteria, was done visually.
\begin{figure}[h]
\includegraphics[width=14cm]{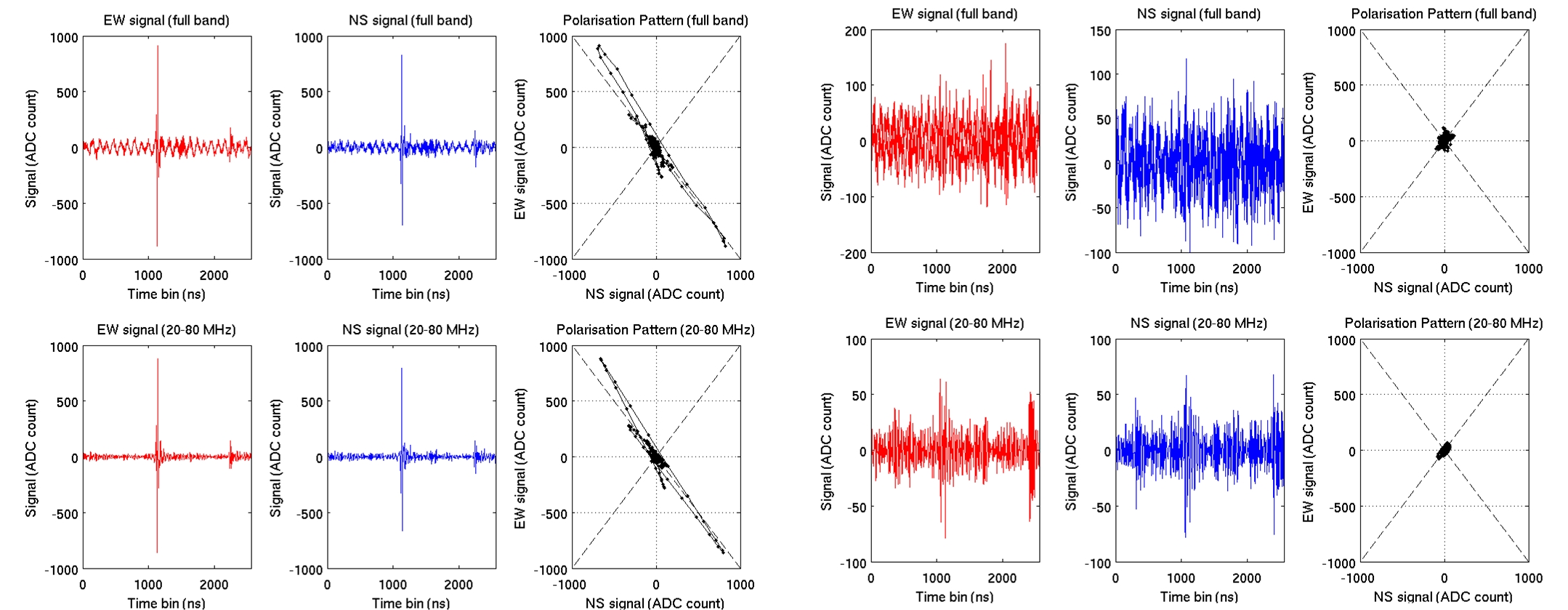}
\centering
\caption{Characteristic example of an event characterized as cosmic \textbf{(left)} and a background event \textbf{(right)}. For each signal the above Figures represent the raw waveform for each pole and the polarization (EW vs NS). The images below represent the corresponding filtered waveforms and the polarization.}
\label{fig:radio_event}
\end{figure}
\begin{table}[h]
\begin{center}
\begin{tabular}{>{\centering\arraybackslash}p{3.3cm}|>{\centering\arraybackslash} p{1.6cm}>{\centering\arraybackslash} p{1.2cm}>{\centering\arraybackslash} p{1.7cm}>{\centering\arraybackslash} p{1.6cm}>{\centering\arraybackslash}p{1.5cm}}
 & Number of events & SNR & Rise time & Polarisation & Total  \\ 
\hline 
Antenna station 1 & 1354 & 76 & 45 & 44 & 44  \\  [0.35cm]
Antenna station 2 & 1292 & 27 & 33 & 23 & 23\\[0.35cm]
\end{tabular} 
\caption{Total number of events, number of events surviving from each independently applied criterion and the number of events passing successfully all three criteria.}
\label{tab:events12}
\end{center}
\end{table} 
Figure \ref{fig:radio_event} presents (on the left) an example of an event that successfully passes all the criteria and finally is characterized as being of cosmic origin, against an event (on the right) characterized as a background event, not passing any criteria. Events in coincidence between stations 1 and 2 are presented in Table \ref{tab:events12}, after passing successfully each criterion separately, and all criteria together. Finally, 44 in station 1 and 23 events in station 2 survive all cuts and are flagged as cosmic origin candidate events.\\
Comparing the absolute times of these 44 and 23 events, using the antenna GPS absolute times and setting the selection limit according to the distance between antennas  1 and 2, as was done earlier, it was found that 15 events, are detected simultaneously from the two RF detectors. A final confirmation of the cosmic origin of these events, passing all three criteria in both antennas, is the comparison of the absolute time difference between the peaks of the signals in each antenna $\Delta t_{ant}$ (absolute GPS time + time bin of peak voltage) and the corresponding time difference $\Delta t_{scint}$ calculated from particle detectors. The quantity $\Delta t_{ant}= t_{1,ant} - t_{2,ant}$ is calculated from the antenna collected data.
\begin{figure}[h]
\includegraphics[width=14cm]{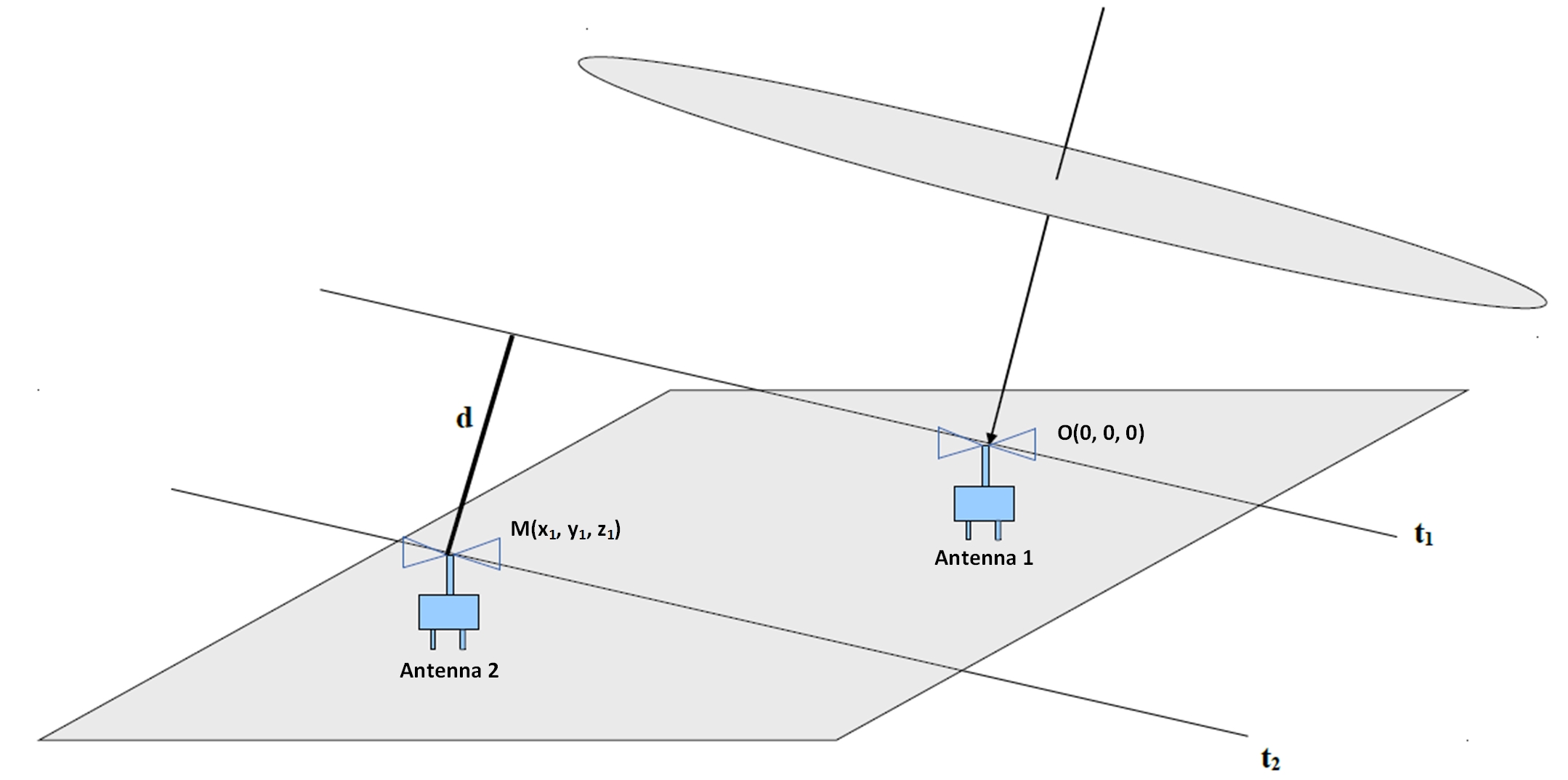}
\centering
\caption{Example of projection of point $M(x_{1},y_{1},z_{1})$, where antenna 2 is located, to the plane perpendicular to the shower axis that passes from the axis center O$( 0, 0, 0)$ where antenna 1 is located.}
\label{fig:ant1_dt}
\end{figure}
The corresponding quantity $\Delta t_{scint}$ was calculated from the zenith and azimuth angles of the incoming primary cosmic ray particle as was estimated using particle detectors timing and the triangulation method. More specifically, the distance $d$ of a point M($x_{1},y_{1},z_{1}$) where antenna 2 is positioned, from a plane perpendicular to the shower axis passing through the point O$(0, 0, 0)$ where antenna 1 is positioned as depicted in Figure \ref{fig:ant1_dt}, is given from Equation \ref{eq:deltat}, 
where $\Delta t_{scint}=d/c$ and c is the speed of light.
\begin{equation}\label{eq:deltat}
d={{|\sin \vartheta \cdot \cos\varphi\cdot x_{1}+\sin\vartheta\cdot\sin\varphi\cdot y_{1}+\cos\vartheta\cdot z_{1}|}   \over
{\sqrt{\left( \sin\vartheta\cdot\cos\varphi\right)^{2}+\left( \sin\vartheta\cdot\sin\varphi\right)^{2}+\cos\vartheta^{2}}}}
\end{equation}
The distribution of the difference of these values $\Delta t =\Delta t_{scint}-\Delta t_{ant}$, is shown  in the left part of Figure \ref{fig:recos_ant1}. 
This distribution is fitted with a Gaussian distribution, with $rms=18.66\,ns$. Taking into account the distance of $162.60\,m$ between the two antennas, with simple calculations this value corresponds to $1.97^{o}$, within the errors estimated from the particle detectors reconstruction. On the contrary, in the right part of Figure \ref{fig:recos_ant1} the distribution of the same difference from a sample of 86 events in double coincidence between the particle detectors stations, but not passing either $SNR$ or rise time criterion. In this case, the distribution (Figure \ref{fig:recos_ant1}, right) has quite greater rms, with value $119.37\,ns$ corresponding to a $12.54^{o}$ estimated error in the direction reconstruction. Thus, there is no doubt for the cosmic origin of the events passing the set criteria, as only in those events the time difference agrees with the corresponding time difference calculated from the shower reconstructed angles $\vartheta$ and $\varphi$, pointing to the direction that the primary particle enters the atmosphere.

\subsection{Noise evaluation}
The mean squared voltage of each of the 11 windows defined in Figure \ref{fig:binning}, for each event passing the $SNR$ criterion is calculated. Then, the mean of these values over every one of the 11 windows are presented in Tables \ref{tab:noise1} and \ref{tab:noise2} for antennas 1 and 2 respectively.
\begin{figure}[h]
\includegraphics[width=14cm]{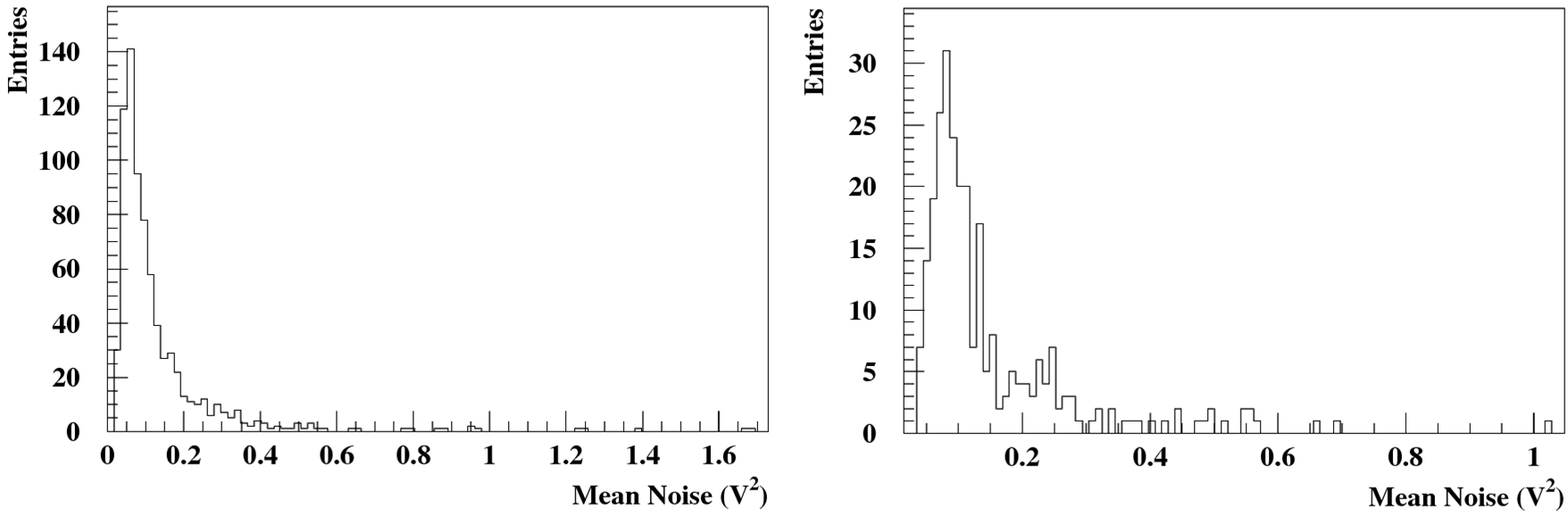}
\centering
\caption{Distribution of mean value of squared voltage for each window containing noise (1 - 5, 7 - 11) for the events of antenna 1 with $SNR>5$ on the \textbf{right} and for the events of antenna 2 with $SNR>8$ on the \textbf{left}.}
\label{fig:noise_12}
\end{figure}
Due to the fulfilment of the $SNR$ criterion, these events do not contain any occasional strong noise sources. By plotting the mean squared voltages for all 10 ``noise" windows for each event, as shown in the left part of Figure \ref{fig:noise_12} for antenna 1 and in the right part of the same Figure  for antenna 2,  the noise mean value  for antenna 1 is calculated to be $\left\langle V^{2}\right\rangle  =0.14\,(V^{2})$ and for antenna 2 $\left\langle V^{2}\right\rangle =0.15\,(V^{2})$ respectively. This value constitutes the background noise to be subtracted during data analysis for the accurate evaluation of physical parameters of the EAS.

\subsection{ Events in coincidence between stations 1 and 3}

The distance between stations 1 and 3 is about $329\,m$. The number of candidate cosmic events detected by station 1 and 3 particle detectors in coincidence has to be smaller in comparison with stations 1 and 2 coincidence events. Table \ref{tab:events13} describes the number of the events passing the criteria in each antenna. In station 1, the $SNR$ cutoff is set again to 5 and the rise time of the cumulative signal to $28\,ns$. The final selection delivers 8 candidate events of cosmic origin.
\begin{table}[h]
\begin{center}
\begin{tabular}{>{\centering\arraybackslash}p{3.3cm}|>{\centering\arraybackslash} p{1.6cm}>{\centering\arraybackslash} p{1.2cm}>{\centering\arraybackslash} p{1.7cm}>{\centering\arraybackslash} p{1.6cm}>{\centering\arraybackslash}p{1.5cm}}
 & Number of events & SNR & Rise time & Polarisation & Total  \\ 
\hline 
Antenna station 1 & 146 & 14 & 18 & 13 & 8  \\  [0.35cm]
Antenna station 3 & 147 & 13 & 6 & 3 & 3\\[0.35cm]
\end{tabular} 
\caption{Total number of events, number of events surviving from each applied criterion and number of events characterized as cosmic, passing successfully all the criteria.}
\label{tab:events13}
\end{center}
\end{table} 
On station 3, the rise time threshold of the cumulative signal is also set to $28\,ns$ and the $SNR$ cutoff value is set to 3. Finally, 3 events passed all criteria and were characterized as candidate events of cosmic origin.
\begin{table}[h]
\begin{center}
\begin{tabular}{>{\centering\arraybackslash}p{1.3cm}|>{\centering\arraybackslash} p{0.7cm}>{\centering\arraybackslash} p{0.7cm}>{\centering\arraybackslash} p{0.7cm}>{\centering\arraybackslash} p{0.7cm}>{\centering\arraybackslash}p{0.7cm}>{\centering\arraybackslash} p{0.7cm}>{\centering\arraybackslash} p{0.7cm}>{\centering\arraybackslash} p{0.7cm}>{\centering\arraybackslash} p{0.7cm}>{\centering\arraybackslash} p{0.7cm}>{\centering\arraybackslash} p{0.7cm}}
Window & 1 & 2 & 3 & 4 & 5 & \textbf{6} & 7 & 8 & 9 & 10 & 11  \\ 
\hline 
$E_i (V^{2})$ & 1.078 & 2.675 & 1.283 & 0.898 & 1.853 & \textbf{7.741} & 2.195 & 0.830 & 0.960 & 1.428 & 0.993  \\  [0.35cm]
\end{tabular} 
\caption{Mean value of squared Voltage, as measured from antenna 3 from cosmic candidate events (after $SNR$ criterion), for events in coincidence between stations 1 and 3, according to the binning of Figure \ref{fig:binning}.}
\label{tab:noise3}
\end{center}
\end{table}
This low number were attributed to constant noise and this fact was quantified in Table \ref{tab:noise3}, produced in the same manner as Table \ref{tab:noise1}  with $SNR>3$. The source of this noise is undoubtedly man made and at the time of writing the present paper it was not localized. For antenna 1, the noise level remained at the same level as in Table \ref{tab:noise1} as expected. The events surviving all criteria in antennas 1 and 3, presented in Table \ref{tab:events13}, are not  in coincidence, meaning it was not possible to detect an EAS with simultaneous RF signal in both antennas 1 and 3. This is possibly due to the large distance between the stations and  the increased noise in antenna 3. 

\section{Coclusions}

A combined scintillator counter and RF  antenna array was setup in the outskirts of the city of Patras, Greece. The initial experimental method recorded information for the particles of the shower front in order to study EAS. Additionally, using a trigger signal from particle detectors, the RF signature of EAS was detected in a noisy city environment. The results show that  the RF component of the EAS indeed can be detected using particle detectors as trigger. More antennas have already been installed to further study and reconstruct EAS using the RF method.  

\section*{Acknowledgments}

This research has been co-financed by the European Union (European Social Fund - ESF) and Greek national funds through the Operational Program ``Education and Lifelong Learning" of the National Strategic Reference Framework (NSRF) - Research Funding Program: ``THALIS - Hellenic Open University - Development and Applications of Novel Instrumentation and Experimental Methods in Astroparticle Physics". We would like to thank Codalema experiment and especially Dr. Lillian Martin for their precious support in every stage of this work, also we want to express our thanks to Codalema researcher Didier Charrier for the filtering code.

\bibliographystyle{unsrt}
\bibliography{antenna_paper}

\end{document}